%% file: dpsfinal.tex
\documentclass{llncs}

\usepackage[dvips]{graphicx}
\usepackage{color}
\usepackage{subfigure}
\usepackage{latexsym}
\usepackage{moreverb}
\usepackage{amssymb}
\usepackage{amsmath}
\usepackage{alltt}
\usepackage{makeidx}
\usepackage{ifthen}

\allowdisplaybreaks

\renewcommand{\color}[2][rgb]{}

\newcommand{\sket}[1]{\ensuremath{\lvert#1\rangle}}
\newcommand{\lket}[1]{\ensuremath{\left\lvert#1\right\rangle}}
\newcommand{\ket}[1]{\if@display\lket{#1}\else\sket{#1}\fi}

\newcommand{\sbra}[1]{\ensuremath{\langle#1\rvert}}
\newcommand{\lbra}[1]{\ensuremath{\left\langle#1\right\rvert}}
\newcommand{\bra}[1]{\if@display\lbra{#1}\else\sbra{#1}\fi}

\newcommand{\sbraket}[2]{\ensuremath{\langle#1\rvert#2\rangle}}
\newcommand{\lbraket}[2]{\ensuremath{\left\langle#1\!\left\rvert\vphantom{#1}#2\right.\!\right\rangle}}
\newcommand{\braket}[2]{\if@display\lbraket{#1}{#2}\else\sbraket{#1}{#2}\fi}

\newcommand{\sproj}[1]{\ensuremath{\lvert#1\rangle\!\langle #1\rvert}}
\newcommand{\lproj}[1]{\ensuremath{\left\lvert#1\right\rangle\!\!\left\langle 
#1\right\rvert}}
\newcommand{\proj}[1]{\if@display\lproj{#1}\else\sproj{#1}\fi}

\newcommand{\hilbert}{\ensuremath{\mathcal{H}}}

\newcommand{\zero}{\ket{0}}

\newcommand{\id}{\ensuremath{\mathbb{I}}}

\DeclareMathOperator{\tr}{tr}
\newcommand{\strace}[2][@]{\ensuremath{\tr\ifthenelse{\equal{#1}{@}}{}{_{#1}}(#2)}}
\newcommand{\ltrace}[2][@]{\ensuremath{\tr\ifthenelse{\equal{#1}{@}}{}{_{#1}}\left(#2\right)}}
\newcommand{\trace}[2][@]{\if@display\ltrace[#1]{#2}\else\strace[#1]{#2}\fi}

\newcommand{\sspan}[1]{\ensuremath{\operatorname{span}(#1)}}
\newcommand{\lspan}[1]{\ensuremath{\operatorname{span}\left(#1\right)\!}}
\newcommand{\vspan}[1]{\if@display\lspan{#1}\else\sspan{#1}\fi}

\newcommand{\plaintext}{\mathcal{P}}
\newcommand{\encryptions}{\mathcal{E}}

\title{On the Key-Uncertainty of Quantum Ciphers and the
Computational Security of One-way
Quantum Transmission}
\author{Ivan Damg{\aa}rd \and Thomas Pedersen\thanks{Part of this 
research was funded by European project PROSECCO.} \and Louis 
Salvail$^\thefootnote{}$}
\institute{BRICS\thanks{
        Funded by the Danish National Research Foundation.},
FICS\thanks{FICS, Foundations in
Cryptography and Security, funded by the Danish Natural Sciences
Research Council.},
              Dept. of Computer Science, University of \AA rhus,\,
              \email{\{ivan|pede|salvail\}@brics.dk}}

\begin{document}

\maketitle

\begin{abstract}
   We consider the scenario where Alice wants to send a secret
   (classical) $n$-bit message to Bob using a classical key, and where
   only one-way transmission from Alice to Bob is possible.  In this
   case, quantum communication cannot help to obtain perfect secrecy
   with key length smaller then $n$.  We study the question of whether
   there might still be fundamental differences between the case where
   quantum as opposed to classical communication is used. In this
   direction, we show that there exist ciphers with perfect security
   producing quantum ciphertext where, even if an adversary knows the
   plaintext and applies an optimal measurement on the ciphertext, his
   Shannon uncertainty about the key used is almost maximal. This is in
   contrast to the classical case where the adversary always learns $n$
   bits of information on the key in a known plaintext attack. We also
   show that there is a limit to how different the classical and
   quantum cases can be: the most probable key, given matching plain-
   and ciphertexts, has the same probability in both the quantum and
   the classical cases.  We suggest an application of our results in
   the case where only a short secret key is available and the message
   is much longer.  Namely, one can use a pseudorandom generator to
   produce from the short key a stream of keys for a quantum cipher,
   using each of them to encrypt an $n$-bit block of the message.  Our
   results suggest that an adversary with bounded resources in a known
   plaintext attack may potentially be in a much harder situation
   against quantum stream-ciphers than against any classical
   stream-cipher with the same parameters.
\end{abstract}

\section{Introduction}
In this paper, we consider the scenario where Alice wants to send a
secret (classical) $n$-bit message to Bob using an $m$-bit classical
shared key, and where only one-way transmission from Alice to Bob is
possible (or at least where interaction is only available with a
prohibitively long delay).  If interaction had been available, we
could have achieved (almost) perfect secrecy using standard quantum
key exchange, even if $m< n$. But with only one-way communication, we
need $m\geq n$ even with quantum communication \cite{Tappetal}.

We study the question of whether there might still be some fundamental
differences between the case where quantum as opposed to classical
communication is used. In this direction, we present two examples of
cryptosystems with perfect security producing $n$-bit quantum
ciphertexts, and with key length $m=n+1$, respectively $m=2n$. We show
that given plaintext and ciphertext, and even when applying an optimal
measurement to the ciphertext, the adversary can learn no more than
$n/2$, respectively $1$ bit of Shannon information on the key. This
should be compared to the fact that for a classical cipher with
perfect security, the adversary always learns $n$ bits of information
on the key. While proving these results, we develop a method which may
be of independent interest, for estimating the maximal amount of
Shannon information that a measurement can extract from a mixture.  We
note that the first example can be implemented without quantum memory,
it only requires technology similar to what is needed for quantum key
exchange, and is therefore within reach of current technology. The
second example can be implemented with a circuit of $O(n^3)$ gates out
of which only $O(n^2)$ are elementary quantum gates.

We also discuss the composition of ciphers, i.e., what happens to the
uncertainty of keys when the same quantum cipher is used to encrypt
several blocks of data using independent keys.  This requires some care,
it is well known that
cryptographic constructions do not always compose nicely in the
quantum case. For composition of our ciphers, however,
we shows that the
adversary's uncertainty about the keys grows linearly  with the
number of blocks encrypted, and in some cases
it can be shown to grow 
exactly as one would expect classically.

On the other hand, we show that there is a limit to how different the
quantum and classical cases can be. Namely, the most probable
key (i.e. the min-entropy of the key),
given matching plain- and ciphertexts, has the same probability
in both cases.

On the technical side, a main observation underlying our results on
Shannon key-uncertainty is that our method for estimating the optimal
measurement w.r.t. Shannon entropy can be combined with known results
on so called entropic uncertainty relations \cite{MU,L,S} and mutually
unbiased bases \cite{WF}. We note that somewhat related techniques are
used in concurrent independent work by DiVincenzo et al.  \cite{DV} to
handle a different, non-cryptographic scenario.

While we believe the above results are interesting, and perhaps even
somewhat surprising from an information theoretic point of view, they
have limited practical significance if perfect security is the goal: a
key must never be reused, and so we do not really have to care whether
the adversary learns information about it when it is used.

However, there is a different potential application of our results to the case
where only a short secret key is available, and where no upper bound
on the message length is known a priori. In such a case,
only computational security is possible and the standard classical
way to encrypt is to use a stream-cipher: using a pseudorandom
generator, we expand the key into a long random looking keystream,
which is then combined with the plaintext to form the ciphertext. The
simplest way of doing such a combination is to take the bit-wise XOR
of key and plaintext streams. In a known plaintext attack, an adversary
will then be able to learn full information on a part of the keystream
and can try to analyze it to find the key or guess other parts
of the keystream better than at random. In general, any cipher with
perfect secrecy, $n$-bit plain- and ciphertext and $m$-bit keys can be
used: we simply take the next $m$ bits from the keystream and use
these as key in the cipher to encrypt the next $n$ bits of the
plaintext.  It is easy to see that for any classical cipher, if the
adversary knows some $n$-bit block of plaintext and also the matching
ciphertext, then he learns $n$ bit of Shannon information on the keystream.

If instead we use quantum communication and one of our quantum ciphers
mentioned above, intuition suggests that an adversary with
limited resources is in a more difficult situation when doing a known
plaintext attack:
if measuring the state representing the ciphertext only reveals a small
amount of information on the corresponding part of the keystream, then the
adversary will need much
more known plaintext than in the classical case  before being able to
cryptanalyze
the keystream.

Care has to be taken in making this statement more precise: our results on key
uncertainty tell us what happens when keys are random, whereas
in this application they are pseudorandom. It is conceivable
that the adversary could design a measurement revealing more information
by exploiting the fact that the keystream is not truly random. This, however,
is equivalent to cryptanalyzing the generator using a {\em quantum}
computation,
and is likely to be technologically much harder than implementing the
quantum ciphers.
In particular, unless the generator is very poorly designed, it will
require keeping
a coherent state much larger than what is required for encryption and
decryption
-- simply because one will need to involve many bits from the keystream
simultaneously in order to distinguish it efficiently from random.
Thus, an adversary limited to
   measurements involving only a {\em small} number of qubits
will simply have to make
many such  measurements, hoping to gather enough classical
information on the keystream
to cryptanalyze it. Our results
apply to this
situation: first, since the adversary makes many measurements, we should worry
about what he learns on average, so Shannon information is the
appropriate measure.
Second, even though the keystream is only pseudorandom, it may be
genuinely random
when considering only a small part of it (see Maurer and Massey \cite{MM}).

In Sect. \ref{app}, we prove a lower bound on the amount of
known plaintext the adversary would need in order to obtain a given
amount of information on
the keystream, for a particular type of keystream generator and
assuming the size of
coherent states the adversary can handle is limited.
We believe that quantum communication
helps even for more general adversaries and generators.
However, quantifying this advantage
is an open  problem. We stress that our
main goal here is merely to point out the potential for improved
security against a bounded
adversary.

\section{Preliminaries}

We assume the reader is familiar with the standard notions of Shannon
entropy $H(\cdot)$ of a probability distribution, conditional entropy, etc.
A related notion that also measures ``how uniform'' a distribution is,
is the so called {\em min-entropy}.  Given a probability distribution
$\{ p_1,...,p_n\}$, the min-entropy is defined as
\begin{equation}
H_{\infty}( p_1,...,p_n ) = -\log_2( max\{ p_1,...,p_n\})
\end{equation}
As usual, $H_{\infty}(X)$ for random variable $X$ is the min-entropy
of its distribution.  Min-entropy is directly related to the ``best
guess'' probability: if we want to guess which value random variable
$X$ will take, the best strategy is to guess at a value with maximal
probability, and then we will be correct with probability
$2^{-H_{\infty}(X)}$.  Given the value of another random variable $Y$,
we can define $H_{\infty}(X| Y=y)$ simply as the min-entropy of the
distribution of $X$ given that $Y=y$, and similarly to Shannon
entropy, we can define $H_{\infty}(X| Y)= \sum_y Pr(Y=y) \cdot
H_{\infty}(X| Y=y)$.

The min-entropy can be thought of as a worst-case measure, which is
more relevant when you have access to only one sample of some random
experiment, whereas Shannon entropy measures what happens on average
over several experiments.  To illustrate the difference, consider the
two distributions $(1/2,1/2)$ and $(1/2,1/4,1/4)$. They both have
min-entropy 1, even though it intuitively seems there should be more
uncertainty in the second case, indeed the Shannon entropies are 1 and
1.5. In fact, we always have $H(X)\geq H_{\infty}(X)$, with equality
if $X$ is uniformly distributed.

\section{Classical Ciphers}

Consider a classical cryptosystem with $n$-bit plain and ciphertexts,
$m$-bit keys and perfect secrecy (assuming, of course, that keys are
used only once). We identify the cryptosystem with its encryption
function $E(\cdot,\cdot)$.  We call this an $(m,n)$-cipher for short.

\begin{definition}
      Consider an $(m,n)$-cipher $E$. We define the {\em Shannon} {\em
        key-un\-cer\-tain\-ty} of $E$ to be the amount of Shannon entropy that
      remains on an $m$-bit key given $n$-bit blocks of plain- and
      ciphertexts, i.e.  $H(K|P,C)$, where $K,P,C$ are random variables
      corresponding to the random choices of key, plaintext and ciphertext
      blocks for $E$, and where the key is uniformly chosen.  The {\em
        min-entropy key-uncertainty} of $E$ is defined similarly, but
      w.r.t. min-entropy, as $H_{\infty}(K|P,C)$.
\end{definition}

    From the definition, it may seem that the key uncertainties depend on
the distribution of the plaintext. Fortunately, this is not the case.
The key-uncertainty in the classical case is easy to compute, using
the following slight generalization of the classical perfect security
result by Shannon:

\begin{proposition}
      Let $E$ be a cipher with perfect security, and with plaintext,
      ciphertext and keyspace $\cal P, C, K$, where $|{\cal P}| = |{\cal
        C}|$. Furthermore, assume that keys are chosen uniformly. For any
      such cipher, it holds that the distribution of the key, given any
      pair of matching ciphertext and plaintext is uniform over a set of
      $|{\cal K}|/|{\cal P}|$ keys.
\end{proposition}
\begin{proof}
      By perfect security, we must have $|{\cal K}|\geq |{\cal P}|$. Now,
      let us represent the cipher in a table as follows: we index rows by
      keys and columns by plaintexts, and we fill each entry in the table
      with the ciphertext resulting from the key and plaintext on the
      relevant row and column. Then, since correct decryption must be
      possible and $|{\cal P}| = |{\cal C}|$, each ciphertext appears
      exactly once in each row. Fix any ciphertext $c$, and let $t_c$ be
      the number of times $c$ appears in, say, the first column. Since the
      probability distribution of the ciphertext must be the same no
      matter the plaintext, $c$ must appear $t_c$ times in every column.
      Since it also appears in every row, it follows that the length of a
      column satisfies $|{\cal K}| = t_c |{\cal P}|$. So $t_c = |{\cal
        K}|/|{\cal P}|$ is the same for every $c$. If we know a matching
      plaintext/ciphertext pair, we are given some $c$ and a column, and
      all we know is that the key corresponds to one of the $t_c$ possible
      rows. The proposition follows.\qed
\end{proof}

\begin{corollary}
      For any classical $(m,n)$-cipher, both the Shannon- and min-en\-tro\-py
      key-uncertainty is $m-n$ bits.
\end{corollary}

This result shows that there is no room for improvement in classical
schemes: the natural constraints on $(m,n)$-ciphers imply that the
key-uncertainty is always the same, once we fix $m$ and $n$. As we
shall see, this is not true for quantum ciphers. Although they cannot
do better in terms of min-entropy key uncertainty, they can when it
comes to Shannon key-uncertainty.

\section{Quantum Ciphers and Min-Entropy Key-Uncertainty}

In this section, we consider quantum ciphers which encrypt classical
messages using classical keys and produce quantum ciphers.

We model both the encryption and decryption processes by unitary
operations on the plaintext possibly together with an ancilla. This
is the same model as used in \cite{Tappetal}, with the restriction
that we only encrypt classical messages.

\begin{definition}[$(m,n)$-quantum cipher]
      A \emph{general $(m,n)$-quantum cipher} is a tuple $(\plaintext,
      \encryptions)$, such that
\begin{itemize}
\item $\plaintext \subseteq \hilbert$ is a finite set of orthonormal
      pure-states (plaintexts) in the Hilbert space $\hilbert$,
and $\|\plaintext\| = N$ and $N= 2^n$.
\item $\encryptions = \{\mathsf{E}_k: \hilbert \to \hilbert |\ k =
      1,\ldots,M \}$ is a set of unitary operators (encryptions), and $M=
      2^m$. Decryption using key $k$ is performed using
      $\mathsf{E}_k^\dag$.
\end{itemize}
And the following properties hold:
\begin{itemize}
\item Key hiding: $(\forall k,k' \in \{1,\ldots,M\})$,
      \begin{equation}
         \sum_{a \in
          \plaintext} \frac{1}{N}
        \mathsf{E}_k\ket{a}\proj{0}\bra{a}\mathsf{E}_k^\dag = \sum_{a \in
          \plaintext}
        \frac{1}{N}\mathsf{E}_{k'}\ket{a}\proj{0}\bra{a}\mathsf{E}_{k'}^\dag.
      \end{equation}
\item Data hiding: $(\forall \ket{a},\ket{b} \in \plaintext)$,
      \begin{equation}
        \sum_{k = 1}^{M}
        \frac{1}{M} \mathsf{E}_k\ket{a}\proj{0}\bra{a}\mathsf{E}_k^\dag =
        \sum_{k = 1}^{M}
        \frac{1}{M}\mathsf{E}_k\ket{b}\proj{0}\bra{b}\mathsf{E}_k^\dag.
      \end{equation}
\end{itemize}
\end{definition}

The key and data hiding properties guarantee that an adversary cannot
gain any information about the key and message respectively when an
arbitrary ciphertext is seen.
In \cite{Tappetal}, it was shown that data hiding implies that $m \geq
n$.

The key hiding property states that an adversary with no information
on the message encrypted expects to see the same ensemble no matter
what key was used. We denote this ensemble
\begin{equation}
      \rho = \sum_{a \in \plaintext} \frac{1}{N}
      \mathsf{E}_k\ket{a}\proj{0}\bra{a}\mathsf{E}_k^\dag,
\end{equation}
for any $k \in \{1,2,\ldots,M\}$. As motivation for the key-hiding property,
we mention that it is always
satisfied if ciphertexts are as short as possible ($dim(\hilbert ) =
2^n$). On the other hand, if the key-hiding property does not hold then
the cipher-state on its own reveals information about the secret-key.
This is certainly an unnecessary weakness that one should avoid when
designing ciphers.

The data hiding property states that the adversary expects to
see the same ensemble no matter what message was encrypted. We denote
this ensemble
\begin{equation}
      \sigma = \sum_{k = 1}^{M} \frac{1}{M}
      \mathsf{E}_k\ket{a}\proj{0}\bra{a}\mathsf{E}_k^\dag,
\end{equation}
for any $a \in \plaintext$. We first prove that $\rho = \sigma$.

\begin{lemma}
      \label{lemma:minentropy}
      $\rho = \sigma$.
\end{lemma}
\begin{proof}
      Define the state
      \begin{equation}
        \xi = \sum_{k = 1}^{M}\sum_{a \in
\plaintext}\frac{1}{MN}\mathsf{E}_k\ket{a}\proj{0}\bra{a}\mathsf{E}_k^\dag.
        \label{eq:xi}
      \end{equation}
      Observe that
      \begin{equation}
        \xi = \sum_{k = 1}^{M}\sum_{a \in
    \plaintext}\frac{1}{MN}\mathsf{E}_k\ket{a}\proj{0}\bra{a}\mathsf{E}_k^\dag
= \sum_{k = 1}^{M}\frac{1}{M}\rho = \rho.
      \end{equation}
      Similarly, when switching the sums in (\ref{eq:xi}), we get $\xi =
      \sigma$. We conclude that $\rho = \sigma$. \qed
\end{proof}

We are now ready to prove that for any $(m,n)$-quantum cipher there
exists a measurement that returns the secret key with probability
$2^{n-m}$ given any plaintext and its associated cipher-state. In
other words and similarly to the classical case, the min-entropy
key-uncertainty of any $(m,n)$-quantum cipher is at most $m-n$.

\begin{theorem}[Min-entropy key uncertainty]
      \label{minentrop}
      Let $(\plaintext, \encryptions)$ be an $(m,n)$-quantum cipher,
      encoding the set $\plaintext$. Then
      \begin{equation}
      (\forall a \in \plaintext)(\exists \text{ POVM }
\{M_i\}_{i=1}^M)(\forall k \in \{1,\ldots,M\})
      [\trace{M_k \mathcal{E}_k(\proj{a})} = 2^{n-m}].
      \end{equation}
\end{theorem}
\begin{proof}
      Let $\ket{a} \in \plaintext$ be given. Consider the set $\mathcal{M}
      = \{M_k = \frac{N}{M}
      \mathsf{E}_k\ket{a}\proj{0}\bra{a}\mathsf{E}_k^\dag\ |\ k=1,\ldots,M
      \}$.
      Lemma \ref{lemma:minentropy} gives
      \begin{equation}
          \sum_{k=1}^M M_k = \sum_{k=1}^M \frac{N}{M}
          \mathsf{E}_k\ket{a}\proj{0}\bra{a}\mathsf{E}_k^\dag= N\sigma = N\rho.
      \end{equation}

      Since the plaintexts are orthogonal quantum states, and since
      unitary operators preserve angles, we have that $N \sum_{a \in
        \plaintext} \frac{1}{N}
      \mathsf{E}_k\ket{a}\proj{0}\bra{a}\mathsf{E}_k^\dag$ is the eigen
      decomposition of $N \rho$, and that $1$ is the only eigenvalue.
      Therefore there exists a positive operator $P$ such that $N \rho + P
      = \id$, and thus
      \begin{equation}
        \sum_{k=1}^M M_k + P = N\rho + P = \id,
      \end{equation}
      and $\mathcal{M} \cup \{P\}$ (and therefore also $\mathcal{M}$) is a
      valid POVM.

      The probability of identifying the key with the measurement
      $\mathcal{M}$ is
      \begin{equation}
        \begin{split}
          \trace{M_k \mathsf{E}_k\ket{a}\proj{0}\bra{a}\mathsf{E}_k^\dag} &=
          \trace{\frac{N}{M}
            \mathsf{E}_k\ket{a}\proj{0}\bra{a}\mathsf{E}_k^\dag
            \mathsf{E}_k\ket{a}\proj{0}\bra{a}\mathsf{E}_k^\dag} \\
          &= \frac{N}{M}
          \trace{\mathsf{E}_k\ket{a}\proj{0}\bra{a}\mathsf{E}_k^\dag} \\
          &= 2^{n-m},
        \end{split}
      \end{equation}
      which proves the theorem.\qed
\end{proof}

\section{Some Example Quantum Ciphers}\label{excip}

In this section, we suggest a general method for designing quantum
ciphers that can do better in terms of Shannon key-uncertainty than
any classical cipher with the same parameters. The properties of our
ciphers are analyzed in the next section.

The first example is extremely simple:
\begin{definition}
      The $H_n$ cipher is an $(n+1,n)$-quantum cipher. Given message
      $b_1,b_2,\ldots,b_n$ and key $c,k_1,\ldots,k_n$, it outputs the
      following $n$ q-bit state as ciphertext:
      \begin{equation}
        (H^{\otimes n})^{c}( X^{k_1}\otimes X^{k_2}\otimes \ldots \otimes
        X^{k_n}\ket{b_1b_2 \ldots b_n} ),
      \end{equation}
      where $X$ is the bit-flip operator and $H$ is the Hadamard
      transform.  That is, we use the last $n$ bits of key as a one-time
      pad, and the first key bit determines whether or not we do a
      Hadamard transform on all $n$ resulting q-bits.
\end{definition}

Decryption is trivial by observing that the operator $ (X^{k_1}\otimes
X^{k_2}\otimes \cdots \otimes X^{k_n})(H^{\otimes n})^{c}$ is the
inverse of the encryption operator. It is also easy to see that the
data hiding property is satisfied: if $c,k_1,\ldots,k_n$ are uniformly
random, then the encryption of any message produces the complete
mixture (in fact this would be the case, already if only
$k_1,\ldots,k_n$ were uniformly random).

This cipher can be described from a more general point of view: let
${\cal B} = \{ B_0,\ldots, B_{2^t-1}\}$ be a set of $2^t$ orthonormal
bases for the Hilbert space of dimension $2^n$. We require that the
bases do not overlap, i.e., no unit vector occurs in more than one
basis. For instance $\cal B$ could consist of the computational basis
and the diagonal basis (i.e. $\{ H^{\otimes n}\ket{x}| x\in \{ 0,1\}^n
\}$). Let $U_i$ be the unitary operator that performs a basis shift
from the computational basis to the basis $B_i$. Finally, let
$[k_1,\ldots,k_t]$ be the number with binary representation
$k_1,\ldots,k_t$.  Then we can define an $(n+t,n)$-cipher $C_{\cal B}$
which on input a key $c_1,\ldots,c_t,k_1,\ldots,k_n$ and a plaintext
$b_1,\ldots,b_n$ outputs
\begin{equation}
      U_{[c_1,\ldots,c_t]}(X^{k_1}\otimes X^{k_2}\otimes \ldots \otimes
      X^{k_n}\ket{b_1b_2 \ldots b_n} ).
\end{equation}
The $H_n$-cipher above is a special case with $U_0 = Id, U_1= H^{\otimes n}$.
Using arguments similar to the above, it is easy to see that

\begin{lemma}
      For any set of orthonormal non-overlapping bases $\cal B$, $C_{\cal
        B}$ is a quantum cipher satisfying the data hiding and unique
      decryption properties.
\end {lemma}

The lemma holds even if $\cal B$ contains only the computational
basis, in which case $C_{\cal B}$ is equivalent to the classical
one-time pad.  The point of having several bases is that if they are
well chosen, this may create additional confusion for the adversary,
so that he will not learn full information on the key, even knowing
the plaintext. We shall see this below.

For now, we note that Wootters and Fields have shown that in a Hilbert
space of dimension $2^n$, there exists $2^n+1$ orthonormal bases that
are {\em mutually unbiased}, i.e., the inner product between any pair
of vectors from different bases has norm $2^{-n/2}$.  Using, say, the
first $2^n$ of these bases, we get immediately from the construction
above a $(2n,n)$ cipher:

\begin{definition}
      The $W_n$-cipher is the cipher $C_{\cal B}$ obtained from the above
      construction when $\cal B$ is the set of $2^n$ mutually
      unbiased bases obtained from \cite{WF}.
\end{definition}

\subsection{Efficient Encoding/Decoding}
\newcommand{\mbf}[1]{\mathbf{#1}}
\newcommand{\bs}[1]{\mbox{$\boldsymbol{#1}$}}

In this section we look at how to implement $W_n$ efficiently. In
\cite{WF}, a construction for $2^n+1$ mutually unbiased bases in the
space of $n$ qubits is given. In the following, we denote by
$v_{\bs{s}}^{(\bs{r})}$ with $\bs{s},\bs{r}\in \{0,1\}^{n}$ the
$\bs{s}$-th vector in the $\bs{r}$-th mutually unbiased basis. We
write $v_{\bs{s}}^{(\bs{r})}$ in the computational basis as,
\begin{equation}\label{v}
      \ket{v_{\bs{s}}^{(\bs{r})}} = \sum_{\bs{l}\in\{0,1\}^n}
      \left(v_{\bs{s}}^{(\bs{r})}\right)_{\bs{l}} \ket{\bs{l}},
\end{equation}
where $\sum_{\bs{l}}|(v_{\bs{s}}^{(\bs{r})})_{\bs{l}}|^2=1$.  Wootters
and Field\cite{WF} have shown that $2^{n}$ mutually unbiased bases are
obtained whenever
\begin{equation}\label{vi}
      \left(v_{\bs{s}}^{(\bs{r})}\right)_{\bs{l}}= \frac{1}{\sqrt{2^{n}}}
      i^{\bs{l}^T(\bs{r}\cdot \bs{\alpha})\bs{l}}(-1)^{\bs{s}\cdot\bs{l}},
\end{equation}
for $\bs{\alpha}$ a vector of $n$ matrices each of dimensions $n\times
n$ with elements in $\{0,1\}$.  The arithmetic in the exponent of $i$
should be carried out over the integers (or equivalently
$\mbox{mod\,}{4}$).  The elements of $\bs{\alpha}$ are defined by
\begin{equation}\label{fifj}
      f_i f_j = \sum_{m=1}^{n} \bs{\alpha}^{(m)}_{i,j}f_m,
\end{equation}
where $\{f_i\}_{i=1}^n$ is a basis for $GF(2^n)$ when seen as a vector
space. Therefore, $\bs{\alpha}$ can be computed on a classical
computer (and on a quantum one) in $O(n^3)$.

Let $c=c_1,\ldots,c_n$ and $k=k_1,\ldots,k_n$ be the $2n$ bits of key
with $c$ defining one out of $2^n$ mutually unbiased basis and $k$
defining the key for the one-time-pad encoding.  The circuit for
encrypting classical message $a$ starts by computing:
\begin{equation}\label{first}
      \ket{\psi_{a}^{k}}  = H^{\otimes n} X^{\otimes
        k} \ket{a}= H^{\otimes n}\ket{a\oplus k}=
      2^{-n/2}\sum_{\bs{l}}(-1)^{(a\oplus k)\cdot \bs{l}}\ket{\bs{l}}.
\end{equation}
The state (\ref{first}) differs from (\ref{v}) only with
respect to the phase factor $i^{\bs{l}^T(\bs{r}\cdot
      \bs{\alpha})\bs{l}}$ in front of each $\ket{\bs{l}}$ with
$\bs{r}=c$.  Transforming (\ref{first}) into (\ref{v}) (i.e.
that is transforming
$\ket{\psi_{a}^{k}}\mapsto\ket{v_{k\oplus
        a}^{(c)}}$) can easily be achieved using a few controlled
operations as described in App. \ref{encrypt}.  The complexity of
the quantum encryption circuit is $O(n^3)$ out of which only $O(n^2)$
are quantum gates.  The decryption circuit is the same as for the
encryption except that it is run in reverse order. A similar
encryption/decryption circuit can easily be implemented for any
$C_{\cal B}$-cipher where ${\cal B}$ is a set of mutually unbiased
bases.

\section{Optimal measurements w.r.t. Shannon Entropy}
Our ultimate goal is to estimate the Shannon key-uncertainty
of an $(m,n)$-quantum cipher, i.e., the amount of entropy that
remains on the key after making an optimal measurement
on a ciphertext where the plaintext is given. But actually, this
scenario is quite general and not tied to the cryptographic application:
what we want to answer is:
given a (pure) state chosen uniformly from a given set of
states, how much Shannon entropy must (at least) remain on the choice of state
after having made a measurement that is optimal w.r.t. minimizing
the entropy?

So what we should
consider is the following experiment: choose a key $k\in \cal K$
uniformly. Encrypt a given plaintext $p$ under key $k$ to
get state $\ket{c_k}$ (we assume here for simplicity that this
is a pure state). Perform some measurement
(that may depend
on $p$) and get outcome $u$. Letting random variables $K,U$
correspond to the choices of key and outcome, we want to
estimate
\begin{equation}
\label{eq0}
H(K|U)= \sum_u Pr(U=u)H(K|U=u).
\end{equation}
Now, $H(K|U= u)$ is simply the Shannon entropy
of the probability distribution
$\{ Pr(K=k|U=u)| k\in {\cal K}\}$. By the standard formula
for conditional probabilities, we have
\begin{equation}
\label{eq1}
Pr(K=k|U=u) = \frac{Pr(U=u|K=k)Pr(K=k)}{Pr(U=u)}.
\end{equation}
Note that neither $Pr(U=u)$, nor $Pr(K=k)$ depend on the
particular value of $k$ (since keys are chosen uniformly).

The measurement in question can be modeled as a POVM, which without
loss of generality can be assumed to contain only elements of the form
$a_u \proj{u}$, i.e., a constant times a projection determined by a
unit vector $\ket{u}$.  This is because the elements of any POVM can
be split in a sum of scaled projections, leading to a measurement with
more outcomes which cannot yield less information than the original
one. It follows immediately that
\begin{equation}
\label{eq2}
Pr(U=u|K=k) = |a_u|^2 |\braket{u}{c_k}|^2.
\end{equation}
Note that also the factor $|a_u|^2$ does not depend on $k$.
Then by (\ref{eq1}) and (\ref{eq2}), we get
\begin{equation}
      1= \sum_{l\in \cal K} Pr(K=l|U=u) =
      \frac{|a_u|^2 Pr(K=l)}{Pr(U=u)}\sum_{l\in \cal K}  |\braket{u}{c_l}|^2.
\end{equation}
Which means that we have
\begin{equation}
      Pr(K=k|U=u)= \frac{ |\braket{u}{c_k}|^2}{\sum_{l\in \cal K}
|\braket{u}{c_l}|^2}.
\end{equation}
In other words, $H(K|U=u)$ can be computed as follows: compute the set
of values $\{ |\braket{u}{c_k}|^2| k\in {\cal K}\}$, multiply by a
normalization factor so that the resulting probabilities sum to 1, and
compute the entropy of the distribution obtained. We call the
resulting entropy $H[\ket{u},S_K]$, where $S_K$ is the set of states
that may occur $\{ \ket{c_k}| k\in {\cal K} \}$. This is to emphasize
that $H[\ket{u},S_K]$ can be computed only from $\ket{u}$ and $S_K$,
we do not need any information about other elements in the
measurement.  From (\ref{eq0}) and $H(K|U=u) = H[\ket{u},S_K]$ follows
immediately
\begin{lemma}
\label{entropy}
With notation as above, we have:
\begin{equation}
H(K|U) \geq min_{\ket{u}}\{ H[\ket{u},S_K]\},
\end{equation}
where $\ket{u}$ runs over all unit vectors in the space
we work in.
\end{lemma}
This bound is not necessarily tight, but it will be, exactly if it is
possible to construct a POVM consisting only of (scaled) projections
$a_u \proj{u}$, that minimize $H[\ket{u},S_K]$.
In general, it may not be easy to solve the minimization problem
suggested by the lemma, particularly if $S_K$ is large and lives in
many dimensions. But in some cases, the problem is tractable, as we shall see.

\section{The Shannon Key-Uncertainty of Quantum Ciphers}
In this section, we study the cipher $C_{\cal B}$ constructed
from a set of $2^t$ orthonormal bases $\cal B$ as defined in Sect. \ref{excip}.
For this, we first need a detour: each basis in our set defines a
projective measurement.  Measuring a state $\ket{u}$ in basis $B_i\in
\cal B$ produces a result, whose probability distribution depends on
$\ket{u}$ and $B_i$. Let $H[\ket{u},B_i]$ be the entropy of this
distribution. We define the Minimal Entropy Sum (MES) of $\cal B$ as
follows:
\begin{equation}
MES({\cal B}) = min_{\ket{u}} \{ \sum_{i=0}^{2^t-1} H[\ket{u},B_i] \},
\end{equation}
where $\ket{u}$ runs over all unit vectors in our space. Lower bounds
on the minimal entropy sum for particular choices of $\cal B$ have
been studied in several papers, under the name of entropic uncertainty
relations \cite{MU,S,L}. This is motivated by the fact that if the sum
is large, then it is impossible to simultaneously have small entropy
on the results of all involved measurements.  One can think of this as
a ``modern'' version of Heisenberg's uncertainty relations. It turns
out that the key uncertainty of $C_{\cal B}$ is directly linked to
$MES({\cal B})$:

\begin{lemma}
\label{uncertainty}
The Shannon key uncertainty of the cipher $C_{\cal B}$ (with $2^t$
bases) is at least
$MES({\cal B})/2^t + t$.
\end{lemma}
\begin{proof}
      We may use Lemma \ref{entropy}, where the set of states $S_K$ in our
      case consists of all basis states belonging to any of the bases in
      $\cal B$.  To compute $H[\ket{u},S_K]$, we need to consider the
      inner products of unit vector $\ket{u}$ with all vectors in $S_K$.
      In our case, this is simply the coordinates of $\ket{u}$ in each of
      the $2^t$ bases, so clearly the norm squares of the inner products
      sum to $2^t$.  Let $z_{ij}$ be the $i$'th vector in the $j$'th basis
      from $\cal B$.  We have,
\begin{equation}
      \begin{split}
        H[\ket{u},S_K] &=  \sum_{j=0}^{2^t-1}\sum_{i=0}^{2^n-1}
        \frac{1}{2^t}|\braket{u}{z_{ij}}|^2
        \log (2^{t}|\braket{u}{z_{ij}}|^{-2})\\
        &= \sum_{j=0}^{2^t-1}\sum_{i=0}^{2^n-1}
\frac{1}{2^t}|\braket{u}{z_{ij}}|^2
        \log (|\braket{u}{z_{ij}}|^{-2}) +
        \sum_{j=0}^{2^t-1}\sum_{i=0}^{2^n-1} \frac{1}{2^t}|\braket{u}{z_{ij}}|^2
        \log (2^{t})\\
        &=\frac{1}{2^t}\sum_{j=0}^{2^t-1}\sum_{i=0}^{2^n-1}
|\braket{u}{z_{ij}}|^2
        \log (|\braket{u}{z_{ij}}|^{-2}) +
        t \frac{1}{2^t}\sum_{j=0}^{2^t-1}\sum_{i=0}^{2^n-1}
|\braket{u}{z_{ij}}|^2\\
        &=\frac{1}{2^t}\sum_{j=0}^{2^t-1} H[\ket{u},B_j] +t \geq
        \frac{1}{2^t} MES({\cal B} ) +t.
        \end{split}
      \end{equation}
The lemma follows.
\qed
\end{proof}
We warn the reader against confusion about the role of $\ket{u}$ and
$\cal B$ at this point. When we estimate the key uncertainty of
$C_{\cal B}$, we are analyzing a POVM, where $\ket{u}$ is one of the
unit vectors defining the POVM.  But when we do the proof of the above
lemma and use the entities $H[\ket{u},B_j]$, we think instead of
$\ket{u}$ as the vector being measured according to basis $B_j$.
There is no contradiction, however, since what matters in both cases
is the inner products of $\ket{u}$ with the vectors in the bases in
$\cal B$.  We are now in a position to give results for our two
concrete ciphers $H_n$ and $W_n$ defined earlier.

\begin{theorem}\label{hn}
The $H_n$-cipher has Shannon key-uncertainty  $n/2+1$ bits.
\end{theorem}
\begin{proof}
      The main result of \cite{MU} states that when $\cal B$ is a set of
      two mutually unbiased bases in a Hilbert space of dimension $2^n$
      then $MES({\cal B})\geq n$. Using Lemma \ref{uncertainty}, it
      follows that $H_n$ has Shannon key-uncertainty at least $n/2+1$.
      Moreover, there exists measurements (i.e. for example the Von
      Neumann measurement in either the rectilinear or Hadamard basis)
      achieving $n/2+1$ bit of Shannon key-uncertainty. The result
      follows.  \qed
\end{proof}

For the case of $W_n$, we can use a result by Larsen\cite{L}.  He
considers the probability distributions induced by measuring a state
$\ket{u}$ in $N+1$ mutually unbiased bases, for a space of dimension
$N$.  Let the set of bases be $B_1,\ldots,B_{N+1}$, and let
$\pi_{\ket{u},i}$ be the collision probability for the $i$'th
distribution, i.e., the sum of the squares of all probabilities in the
distribution.  Then Larsen's result (actually a special case of it)
says that
\begin{equation}
\sum_{i=1}^{N+1} \pi_{\ket{u},i} = 2
\end{equation}
In our case, $N=2^n$. However, to apply this to our cipher $W_n$, we
would like to look at a set of only $2^n$ bases and we want a bound on
the sum of the entropies $H[\ket{u},B_i]$ and not the sum of the
collision probabilities. This can be solved following a line of
arguments from S\'anchez-Ruiz\cite{S}. Using Jensen's inequality, we
obtain the following:
\begin{equation}\label{sanchez}
      \begin{split}
        \sum_{i=1}^N H[\ket{u},B_i] & \geq  - \sum_{i=1}^N \log \pi
        _{\ket{u},i}\\
        &\geq   -N \log \left( \frac{1}{N}\sum_{i=1}^N  \pi_{\ket{u},i}
      \right) \\
      & = -N \log \left( \frac{1}{N} \left( -\pi_{\ket{u},N+1} +
      \sum_{i=1}^{N+1}  \pi_{\ket{u},i}  \right)\right) \\
      & = N \log \left( \frac{N}{2-\pi_{\ket{u},N+1}} \right) \geq N \log
      \left( \frac{N}{2- 1/N} \right).
\end{split}
\end{equation}
Together with Lemma \ref{uncertainty}, we get:
\begin{theorem}\label{wn}
The $W_n$-cipher has Shannon key-uncertainty greater than $2n-1$ bits.
\end{theorem}
Unlike for $H_n$ (i.e. Theorem \ref{hn}), Theorem \ref{wn} only
provides a lower bound for the key uncertainty
of $W_n$.

Let ${\cal B}$ be any set of $2^t$ mutually unbiased bases living
in a Hilbert space of dimension $2^n$. 
The largest value we could hope for $MES({\cal B})$
is $(2^t-1)n$ bits, since this value is 
exactly matched
when  the state measured is a state that belongs to a basis in ${\cal B}$.
It is natural to define $\Delta(n,t)$ as the distance between $MES({\cal B})$
and the the maximum possible value:
%
%
$$
\Delta(n,t) = (2^t-1)n - 
MES({\cal B}).
$$
Given what we know already,
it seems reasonable to conjecture 
that $\Delta(n,t)$ is, in some sense, small:
we know that 
$\Delta(n,1)=0$ and also that 
$\Delta(n,n) \leq  (2^n-1)n - 
2^n(n-1) = 2^n-n$. Let us consider the following
conjecture:
\begin{conjecture}
\label{conj1}
For any set $\cal B$ containing $2^n$ mutually unbiased bases in a Hilbert 
space
of dimension $2^n$, it holds that
$\frac{\Delta(n,n)}{2^n}\in o(1)$ (i.e. note that we 
know the fraction is strictly smaller than 1).
\end{conjecture}
In this case, 
we easily conclude that cipher $W_n$ has almost full Shannon key-uncertainty:
\begin{lemma}
\label{lemun}
     Under Conjecture \ref{conj1}, $W_n$ has Shannon key-uncertainty at least
    $2n - o(1)$ bits.
\end{lemma}
\begin{proof} From Lemma \ref{uncertainty}, the Shannon key-uncertainty of 
$W_n$ is at least $n+ MES({\cal B})/2^n$. Conjecture \ref{conj1}
leads to
$MES({\cal B})/2^n= ((2^n-1)n-\Delta(n,n))/2^{n}= n-o(1)$.
The result follows.
\qed
\end{proof}

The $H_n$ and $W_n$-ciphers represent two extremes, using the minimal
non-trivial number of bases, respectively as many of the known
mutually unbiased bases as we can address with an integral number of
key bits. It is not hard to define example ciphers that are ``in
between'' and prove results on their key-uncertainty using the same
techniques as for $W_n$.  However, what can be derived from
Larsen's result using the above line of argument (i.e. Equation \ref{sanchez}) 
becomes weaker as one
considers a smaller number of bases.

\section{Composing Ciphers}
What happens to the key uncertainty if we use a quantum cipher twice
to encrypt two
plaintext blocks, using independently chosen keys? Intuition based on
classical behavior
suggests that the key uncertainty should now be twice that of a
single application of the
cipher, since the keys are independent. But in the quantum case, this
requires proof:
the adversary will be measuring a product state composed of 
of two ciphertext
blocks. If the adversary was to measure each block individually then
clearly the key uncertainty would be twice the
key uncertainty of a single block. However, coherent measurements involving both
blocks simultaneously may provide more information on the key
than what is achievable by measuring
the blocks individually.

In the following, we consider composition of the cipher $C_{\cal B}$
with itself,
where  $\cal B$ consists of $2^t$  bases for a space of dimension $2^n$.
This is a $(2(t+n),2n)$-cipher which we call $C_{\cal B}^2$.  Say $\cal
B$ consists of the bases
${\cal B} = \{ B_0,....,B_{2^t-1}\}$. Let us consider the tensor
product of two Hilbert spaces
of dimension $2^n$ each. Then $B_i\otimes B_j$ denotes the basis of
this tensor product
space that one obtains by taking all pairwise tensor products of the
$2^n$ basis vectors in
each of $B_i$ and $B_j$. We will let ${\cal B}\otimes {\cal B}$
denote the set of all
$2^{2t}$ bases that can be formed this way. Since each such basis
consists of $2^{2n}$ basis
vectors, ${\cal B}\otimes {\cal B}$ can also be thought of as a
collection of $2^{2t+2n}$
pure states.

On the adversary's point of view, determining the two $t+n$-bit keys from two
ciphertext blocks is equivalent to the following experiment: choose
uniformly a state in ${\cal B}\otimes {\cal B}$, now
the adversary wants to
make a measurement that minimizes the uncertainty about the state that was picked.

To study this question, we split  ${\cal B}\otimes {\cal B}$ in subsets: let
${\cal B}_i$ be the set of $2^t$ bases defined by
\begin{equation}\label{bi}
{\cal B}_i = \{ B_j\otimes B_{j+i \ \bmod 2^t} |\   j=0,1,...,2^t-1\}
\end{equation}
It is now easy to see that ${\cal B}\otimes {\cal B}$ is the disjoint
union of the
${\cal B}_i$'s, for $i=0,1,...,2^t-1$.

Now, the choice of a state in ${\cal B}\otimes {\cal B}$ can be
rephrased as follows: choose $i$
uniformly from $[0..2^t-1]$, and then choose a state uniformly from
${\cal B}_i$.
Let $I,J$ be random variables representing these choices, and let $U$
be the random
variable representing the adversary's measurement result. 
Standard properties of Shannon entropy give:
\[
H(I,J |\ U) = H(I|\ U) + H(J|\ I,U).
\]

It is straightforward to see that a uniform mixture over all $2^{t+2n}$
states in ${\cal B}_i$ is in fact the complete mixture, and so has
the same density matrix for
any $i$, hence no measurement can reveal information on $I$ and we
have $H(I|\ U) = t$.
We define $M_2({\cal B})=  min_{i}\{ MES({\cal B}_i)\}$. Then,
using exactly the same line of argument as for Lemma
\ref{uncertainty}, one finds that
for each particular value of $i$, we have $H(J|\ I=i ,U) \geq t +
MES({\cal B}_i)/2^t$ and
hence
$H(J|\ I,U) \geq t + M_2({\cal B})/2^t$. Putting things together gives,
\begin{lemma}
$C_{\cal B}^2$ has Shannon key-uncertainty at least $2t + M_2({\cal B})/2^t$.
\end{lemma}
Considering composition of $C_{\cal B}$ $v$ times with itself,
denoted $C_{\cal B}^v$, the
techniques above extend in a straightforward way. In particular, we
end up defining a minimum
$M_v({\cal B})$ over entropy sums for a generalization of the ${\cal
B}_i$'s. This leads to,
\begin{lemma}\label{cvb}
$C_{\cal B}^v$ has Shannon key-uncertainty at least $vt + M_v({\cal B})/2^t$.
\end{lemma}

Note that by the construction defined in (\ref{bi}), 
each ${\cal B}_i$ is a set of mutually unbiased bases, 
and this holds also for any of the $v$-wise generalizations.
In the special case of $H_n$, we have
$t=1$, and each ${\cal B}_i$ (as well as its
$v$-wise generalization) contains 2 mutually unbiased bases. 
Lemma \ref{cvb} together with the
result of
\cite{MU} (i.e. which in our notation reads $M_v({\cal B})=vn$) immediately implies,
\begin{theorem}
The cipher $H_n^v$ has Shannon key uncertainty $v(n/2 +1)$ bits.
\end{theorem}

We do not know of any strong results on the minimal  entropy sum for
any set of mutually
unbiased bases except when its cardinality is 2\cite{MU} or is
close to the dimension of the
space\cite{L,S}. Therefore, we cannot prove a good lower bound 
on the Shannon key-uncertainty for the composition of
$W_n$. Already for $W_n^2$, we need to consider a set of $2^n$
mutually unbiased
bases living in a space of dimension $2^{2n}$. Using the notation of the 
previous
section, we need to bound $\Delta(2n,n)$, or more generally 
$\Delta(vn,n)$.

While $\Delta(vn,n)=0$ may be too much to hope for, it seems
reasonable to conjecture a result similar to the one we know for
$\Delta(n,n)$:
\begin{conjecture}
\label{conj2}
For any set $\cal B$  of $2^n$ mutually unbiased bases living in a Hilbert
space of dimension $2^{vn}$, it holds that $\Delta(vn,n) \leq 2^n
-vn$.
\end{conjecture}
We then have,
\begin{lemma}
  \label{lemmaWn}
  Under Conjecture \ref{conj2}, $W_n^v$ has Shannon key-uncertainty at
  least $2vn - 1$ bits.
\end{lemma}

\section{Application to Stream-Ciphers}
\label{app}

We can use the quantum ciphers we just described to build a
(computationally secure) quantum stream-cipher using a short key $K$
of length independent from the message length. In fact, any
$(m,n)$-cipher and classical pseudorandom generator can be used: we
seed the generator with key $K$, and use its output as a keystream.
To encrypt, we simply take the next $m$ bits from the keystream and
use these as key in the cipher to encrypt the next $n$ bits of the
plaintext.

  
Since an $(m,n)$-cipher has perfect security, this construction would
have perfect security as well if the keystream was genuinely random.
By a standard reduction, this implies that breaking it is at least as
hard as distinguishing the output of the generator from a truly random
string.

All this is true whether we use a classical or an $(m,n)$-quantum
cipher. However, by our results on Shannon key-uncertainty, the
adversary is in a potentially much harder situation in the quantum
case. For intuition on this, we refer to the discussion in the
introduction.  As a more concrete illustration, we consider the
following scenario:
\begin{enumerate}
\item We have a pseudorandom generator $G$, expanding a $k$-bit seed
  $K$ into an $N$-bit sequence $G(K)$. Furthermore, any subset
  containing at most $e$ bits of $G(K)$ is uniformly random.  Finally,
  no polynomial time (in $k$) classical algorithm can with
  non-negligible advantage distinguish $G(K)$ from a truly random
  sequence when given any piece of data that is generated from $G(K)$
  and contains at most $t$ bits of Shannon information on $G(K)$.
  Both $e$ and $t$ are assumed to be polynomial in $k$.\label{a1}
\item Coherent measurements simultaneously involving $\mu$ qubits or
  more are not possible to implement in practice.  However, technology
  has advanced so that the $W_{n}$-cipher can be implemented for some
  $n << \mu$.\label{a2}
\item We will consider an adversary that first obtains some amount of
  known plaintext. Given the plaintext, he decides on a number of
  complete measurements that he executes on parts of the ciphertext
  (under the constraints of assumption 2). For simplicity we assume
  that each measurement involves an integral number of $n$-bit
  ciphertext blocks.\footnote{This assumption can be dropped so that
    we can still prove Lemma \ref{awn} using a more complicated
    argument and provided the local randomness of the generator is
    expanded from $e$ to $n^2e$} Finally he executes any polynomial
  time classical algorithm to analyze the results.\label{a3}
\end{enumerate}
The first assumption can be justified using a result by Maurer and
Massey \cite{MM} on locally random pseudorandom generators.  Their
result asserts that there exists pseudorandom generators satisfying
the assumption that any $e$ bits are genuinely random, provided $e\leq
k/\log_2 N$.  Their generators may not behave well against attacks
having access to more than $e$ bits of the sequence, but one can
always xor the output from their generator with the output of a more
conventional one using an independent key. This will preserve the
local randomness.


Note that the size of $k$ does not influence the size of the quantum
computer required for the honest party to encrypt or decrypt.  The
third assumption essentially says that we do not expect that results
of (incomplete) measurements obtained on one part of the ciphertext
will help significantly in designing measurements on other parts. This
is justified, as long as not too many measurements are performed: as
long as results from previous measurements contain less than $t$ bits
of information on the keystream, then by assumption \ref{a1}, these
results might (from the adversary's point of view) as well have been
generated from measuring a random source, and so they do not help in
designing the next measurement. This assumption can therefore be
dropped in a more careful analysis since it esssentially follows from
assumptions \ref{a1} and \ref{a2}.  For simplicity, we choose to make
it explicit.

\begin{lemma}\label{awn}
  Assume we apply the $W_n$-cipher for stream encryption using a
  pseudorandom generator and with an adversary as defined by
  assumptions \ref{a1},\ref{a2}, and \ref{a3} above.  Suppose we
  choose $e=2\mu$ and $k\geq 2\mu \log_2{N}$.  Then, assuming
  Conjecture \ref{conj2}, the adversary will need to obtain $tn$ bits
  of known plaintext, in order to distinguish the case of a real
  encryption from the case where the keystream is random.
\end{lemma}
\begin{proof}
  Assume the PRG satisfies assumption \ref{a1} which is possible since
  $k\geq e\log_2{N}$.  By assumption \ref{a2}, any attack that
  measures several blocks of ciphertext in one coherent measurement
  can handle at most $\mu=e/2$ qubits at any one time. By
  construction, this ciphertext was created using less than $e$ bits
  of the keystream, which is random by assumption \ref{a1}.
  Therefore, the measurement will give the same result as when
  attacking the composition $W_{n}^{v/n}$ since the measurement
  involves $v\leq \mu$ qubits (since different blocks of the keystream
  are independent if the stream is truly random) and by assumption
  \ref{a3}.  Hence, by Lemma \ref{lemmaWn} and under Conjecture
  \ref{conj2}, the adversary learns less than $1$ bit of information
  on the key stream from each measurement.  Now, if the adversary has
  $T$ bits of known plaintext, and hence measures $T$ ciphertext bits,
  the maximal number of measurements that can take place is $T/n$ so
  he needs to have $T/n > t$ in order for the classical distinguisher
  to work, by assumption \ref{a1}.  The lemma follows.\qed
\end{proof}
This lemma essentially says that for a generator with the right
properties, and for an adversary constrained as we have assumed,
quantum communication allows using the generator securely to encrypt
$tn$ bits, rather than the $t$ bits we would have in the classical
case. Depending on how close the actual key uncertainty of
compositions of $W_n$ is to the maximal value, the number of required
plaintext bits can be much larger. The best we can hope for would be
if $\Delta(vn,n)=0$ for all $n,v$, in which case the adversary would
need $t2^n$ plaintext bits.
 
A similar result can be shown without assuming any conjecture for the
$H_n$ cipher. In this case, we gain essentially a factor
2 in plaintext size over the classical case.
  
Of course, these results do not allow to handle adversaries as general
as we would like, our constraints are different from
just assuming the adversary is quantum polynomial time. Nevertheless,
we believe that the scenario we have described can be reasonable with
technology available in the foreseeable future. Moreover, it seems to
us that quantum communication should help even for more general
adversaries and generators. Quantifying this advantage is an open
problem.

\section{Conclusion and Open Problems}

We have seen that, despite the fact that quantum communication cannot
help to provide perfect security with shorter keys when only one-way
communication is used, there are fundamental differences between
classical and quantum ciphers with perfect security, in particular the
Shannon key uncertainty can be much larger in the quantum case.
However, the min-entropy key-uncertainty is the same in the two cases.
It is an open question whether encryption performed by general quantum
operations allows for quantum ciphers to have more min-entropy
key-uncertainty than classical ones.

We have also seen an application of the results on Shannon key
uncertainty to some example quantum ciphers that could be used to
construct a quantum stream-cipher where, under a known plaintext
attack, a resource-bounded adversary would be in a potentially much
worse situation than with any classical stream-cipher with the same
parameters.

For the ciphers we presented, the Shannon key-uncertainty is known
exactly for the $H_n$-cipher but not for the $W_n$-cipher.  It is an
interesting open question to determine it. More generally, are
Conjectures \ref{conj1} and \ref{conj2} true?

\section*{Acknowledgements}
We are grateful to Renato Renner for pointing out a mistake in the
proof of the Shannon key-uncertainty for the 
composition of cipher $C_{\cal B}$ appearing in the proceedings of
Eurocrypt 2004 (i.e. the proof of Theorem $4$ in\cite{euro04} is wrong!).


\appendix

\section{Encryption Circuit for the $W_n$-Cipher}\label{encrypt}
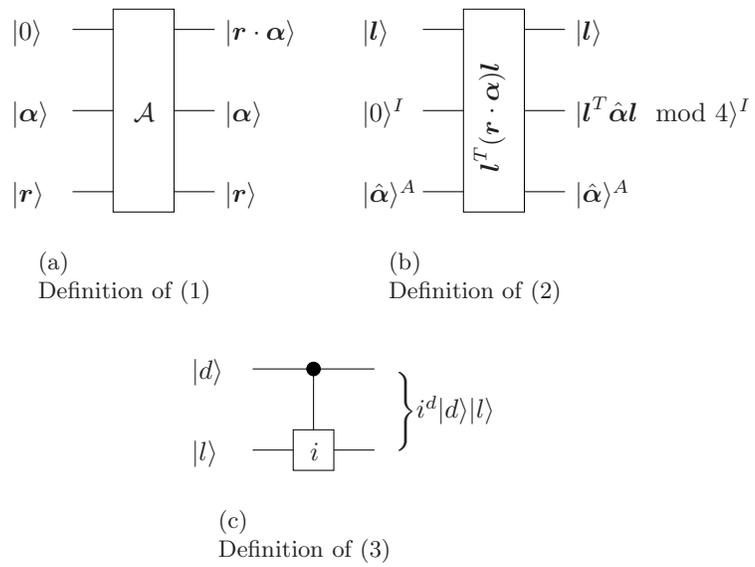
\begin{figure}
        \begin{center}
          \subfigure[\mbox{Definition of (1)}]{\input{unitary-a.pstex_t}}
          \hspace{1.5cm}
          \subfigure[\mbox{Definition of 
(2)}]{\input{unitary-exponent.pstex_t}}\\
          \subfigure[\mbox{Definition of (3)}]{\input{unitary-i.pstex_t}}
        \end{center}
\caption{Sub-circuits to the encryption circuit of Fig. 
\ref{codingcircuit}.}\label{details}
\end{figure}

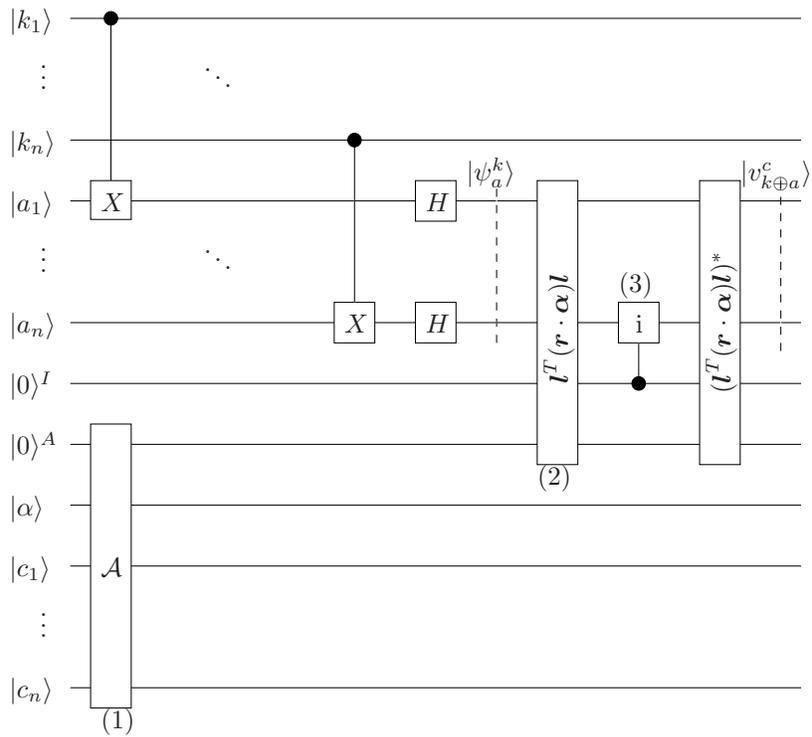
\begin{figure}[h]
\begin{center}
\input{coding-circuit.pstex_t}
\caption{Encoding circuit for cipher $W_n$.}\label{codingcircuit}
\end{center}
\end{figure}

The circuit depicted in Fig. \ref{codingcircuit}
implements the encryption of any plaintext
$a=a_1,\ldots,a_n\in\{0,1\}^n$
according the secret key $(c,k) \in \{0,1\}^{2n}$.
It uses three sub-circuits $(1),(2)$, and $(3)$
as defined in Fig. \ref{details}.
\noindent

${\cal A}$, given $c$ and $\bs{\alpha}$, produces the matrix $c\cdot
\bs{\alpha}$ in the register denoted $A$.  Notice that circuit ${\cal
   A}$ is a classical circuit. It can be implemented with $O(n^3)$
classical gates.  The sub-circuit (2) accepts as input
$\hat{\bs{\alpha}}=c\cdot \bs{\alpha}$ together with $\bs{l}$,
computes $d=\bs{l}^{T}\hat{\bs{\alpha}}\bs{l}\in [0,\ldots,3]$, and
stores the result in a 2-qubit register $I$. In (3), an overall phase
factor $i^{d}$ is computed in front of the computational basis element
$\ket{\bs{l}}$. The last gates allow to reset registers $I$ and $A$
making sure registers containing the encrypted data are separable from
the other registers.  It is straightforward to verify that registers
initially in state $\ket{a_1}\otimes\ldots\otimes\ket{a_n}$ ends up in
state $\ket{v_{k\oplus a}^{(c)}}$ as required.  The overall complexity
is $O(n^2)$ quantum gates since (3) requires only $O(n^2)$ {\sc
   cnot}'s which is of the same complexity as super-gate (2). In
conclusion, the total numbers of gates is $O(n^3)$ out of which
$O(n^2)$ are quantum.

\end{document}

%% file: unitary-a.pstex_t
\begin{picture}(0,0)%
\includegraphics{unitary-a.pstex}%
\end{picture}%
\setlength{\unitlength}{3355sp}%
\begingroup\makeatletter\ifx\SetFigFont\undefined%
\gdef\SetFigFont#1#2#3#4#5{%
  \reset@font\fontsize{#1}{#2pt}%
  \fontfamily{#3}\fontseries{#4}\fontshape{#5}%
  \selectfont}%
\fi\endgroup%
\begin{picture}(1575,1524)(1051,-1123)
\put(1051,164){\makebox(0,0)[lb]{\smash{\SetFigFont{10}{12.0}{\familydefault}{\mddefault}{\updefault}{\color[rgb]{0,0,0}\zero}%
}}}
\put(1051,-436){\makebox(0,0)[lb]{\smash{\SetFigFont{10}{12.0}{\familydefault}{\mddefault}{\updefault}{\color[rgb]{0,0,0}\ket{\bs{\alpha}}}%
}}}
\put(1051,-1036){\makebox(0,0)[lb]{\smash{\SetFigFont{10}{12.0}{\familydefault}{\mddefault}{\updefault}{\color[rgb]{0,0,0}\ket{\bs{r}}}%
}}}
\put(1951,-436){\makebox(0,0)[lb]{\smash{\SetFigFont{10}{12.0}{\familydefault}{\mddefault}{\updefault}{\color[rgb]{0,0,0}${\cal A}$}%
}}}
\put(2626,164){\makebox(0,0)[lb]{\smash{\SetFigFont{10}{12.0}{\familydefault}{\mddefault}{\updefault}{\color[rgb]{0,0,0}\ket{\bs{r}\cdot\bs{\alpha}}}%
}}}
\put(2626,-436){\makebox(0,0)[lb]{\smash{\SetFigFont{10}{12.0}{\familydefault}{\mddefault}{\updefault}{\color[rgb]{0,0,0}\ket{\bs{\alpha}}}%
}}}
\put(2626,-1036){\makebox(0,0)[lb]{\smash{\SetFigFont{10}{12.0}{\familydefault}{\mddefault}{\updefault}{\color[rgb]{0,0,0}\ket{\bs{r}}}%
}}}
\end{picture}

%% file: unitary-exponent.pstex_t
\begin{picture}(0,0)%
\includegraphics{unitary-exponent.pstex}%
\end{picture}%
\setlength{\unitlength}{3355sp}%
\begingroup\makeatletter\ifx\SetFigFont\undefined%
\gdef\SetFigFont#1#2#3#4#5{%
  \reset@font\fontsize{#1}{#2pt}%
  \fontfamily{#3}\fontseries{#4}\fontshape{#5}%
  \selectfont}%
\fi\endgroup%
\begin{picture}(1575,1524)(451,-1123)
\put(451,164){\makebox(0,0)[lb]{\smash{\SetFigFont{10}{12.0}{\familydefault}{\mddefault}{\updefault}{\color[rgb]{0,0,0}\ket{\bs{l}}}%
}}}
\put(451,-436){\makebox(0,0)[lb]{\smash{\SetFigFont{10}{12.0}{\familydefault}{\mddefault}{\updefault}{\color[rgb]{0,0,0}$\zero^I$}%
}}}
\put(451,-1036){\makebox(0,0)[lb]{\smash{\SetFigFont{10}{12.0}{\familydefault}{\mddefault}{\updefault}{\color[rgb]{0,0,0}$\ket{\hat{\bs{\alpha}}}^A$}%
}}}
\put(2026,164){\makebox(0,0)[lb]{\smash{\SetFigFont{10}{12.0}{\familydefault}{\mddefault}{\updefault}{\color[rgb]{0,0,0}\ket{\bs{l}}}%
}}}
\put(2026,-1036){\makebox(0,0)[lb]{\smash{\SetFigFont{10}{12.0}{\familydefault}{\mddefault}{\updefault}{\color[rgb]{0,0,0}$\ket{\hat{\bs{\alpha}}}^A$}%
}}}
\put(2026,-436){\makebox(0,0)[lb]{\smash{\SetFigFont{10}{12.0}{\familydefault}{\mddefault}{\updefault}{\color[rgb]{0,0,0}$\ket{\bs{l}^T\hat{\bs{\alpha}}\bs{l} \mod 4}^I$}%
}}}
\put(1461,-811){\rotatebox{90.0}{\makebox(0,0)[lb]{\smash{\SetFigFont{10}{12.0}{\familydefault}{\mddefault}{\updefault}{\color[rgb]{0,0,0}$\bs{l}^T(\bs{r}\cdot\bs{\alpha})\bs{l}$}%
}}}}
\end{picture}

%% file: unitary-i.pstex_t
\begin{picture}(0,0)%
\includegraphics{unitary-i.pstex}%
\end{picture}%
\setlength{\unitlength}{3355sp}%
\begingroup\makeatletter\ifx\SetFigFont\undefined%
\gdef\SetFigFont#1#2#3#4#5{%
  \reset@font\fontsize{#1}{#2pt}%
  \fontfamily{#3}\fontseries{#4}\fontshape{#5}%
  \selectfont}%
\fi\endgroup%
\begin{picture}(1500,831)(901,-823)
\put(901,-736){\makebox(0,0)[lb]{\smash{\SetFigFont{10}{12.0}{\familydefault}{\mddefault}{\updefault}{\color[rgb]{0,0,0}\ket{l}}%
}}}
\put(1776,-736){\makebox(0,0)[lb]{\smash{\SetFigFont{10}{12.0}{\familydefault}{\mddefault}{\updefault}{\color[rgb]{0,0,0}$i$}%
}}}
\put(901,-136){\makebox(0,0)[lb]{\smash{\SetFigFont{10}{12.0}{\familydefault}{\mddefault}{\updefault}{\color[rgb]{0,0,0}\ket{d}}%
}}}
\put(2401,-421){\makebox(0,0)[lb]{\smash{\SetFigFont{10}{12.0}{\familydefault}{\mddefault}{\updefault}{\color[rgb]{0,0,0}$\Biggl\}i^d\ket{d}\ket{l}$}%
}}}
\end{picture}

%% file: coding-circuit.pstex_t
\begin{picture}(0,0)%
\includegraphics{coding-circuit.pstex}%
\end{picture}%
\setlength{\unitlength}{3355sp}%
\begingroup\makeatletter\ifx\SetFigFont\undefined%
\gdef\SetFigFont#1#2#3#4#5{%
  \reset@font\fontsize{#1}{#2pt}%
  \fontfamily{#3}\fontseries{#4}\fontshape{#5}%
  \selectfont}%
\fi\endgroup%
\begin{picture}(5862,5373)(1651,-5065)
\put(5551,-3211){\makebox(0,0)[lb]{\smash{\SetFigFont{10}{12.0}{\familydefault}{\mddefault}{\updefault}{\color[rgb]{0,0,0}(2)}%
}}}
\put(6151,-1786){\makebox(0,0)[lb]{\smash{\SetFigFont{10}{12.0}{\familydefault}{\mddefault}{\updefault}{\color[rgb]{0,0,0}(3)}%
}}}
\put(2326,-5011){\makebox(0,0)[lb]{\smash{\SetFigFont{10}{12.0}{\familydefault}{\mddefault}{\updefault}{\color[rgb]{0,0,0}(1)}%
}}}
\put(3076,-1636){\makebox(0,0)[lb]{\smash{\SetFigFont{10}{12.0}{\familydefault}{\mddefault}{\updefault}{\color[rgb]{0,0,0}$\ddots$}%
}}}
\put(3076,-286){\makebox(0,0)[lb]{\smash{\SetFigFont{10}{12.0}{\familydefault}{\mddefault}{\updefault}{\color[rgb]{0,0,0}$\ddots$}%
}}}
\put(5026,-961){\makebox(0,0)[lb]{\smash{\SetFigFont{10}{12.0}{\familydefault}{\mddefault}{\updefault}{\color[rgb]{0,0,0}\ket{\psi_a^k}}%
}}}
\put(7051,-961){\makebox(0,0)[lb]{\smash{\SetFigFont{10}{12.0}{\familydefault}{\mddefault}{\updefault}{\color[rgb]{0,0,0}\ket{v_{k \oplus a}^c}}%
}}}
\put(6276,-2086){\makebox(0,0)[lb]{\smash{\SetFigFont{10}{12.0}{\familydefault}{\mddefault}{\updefault}{\color[rgb]{0,0,0}i}%
}}}
\put(1651,-2986){\makebox(0,0)[lb]{\smash{\SetFigFont{10}{12.0}{\familydefault}{\mddefault}{\updefault}{\color[rgb]{0,0,0}$\zero^A$}%
}}}
\put(1651,-2536){\makebox(0,0)[lb]{\smash{\SetFigFont{10}{12.0}{\familydefault}{\mddefault}{\updefault}{\color[rgb]{0,0,0}$\zero^I$}%
}}}
\put(1651,-2086){\makebox(0,0)[lb]{\smash{\SetFigFont{10}{12.0}{\familydefault}{\mddefault}{\updefault}{\color[rgb]{0,0,0}\ket{a_n}}%
}}}
\put(1651,-3436){\makebox(0,0)[lb]{\smash{\SetFigFont{10}{12.0}{\familydefault}{\mddefault}{\updefault}{\color[rgb]{0,0,0}\ket{\alpha}}%
}}}
\put(1876,-4336){\makebox(0,0)[lb]{\smash{\SetFigFont{10}{12.0}{\familydefault}{\mddefault}{\updefault}{\color[rgb]{0,0,0}$\vdots$}%
}}}
\put(1651,-3886){\makebox(0,0)[lb]{\smash{\SetFigFont{10}{12.0}{\familydefault}{\mddefault}{\updefault}{\color[rgb]{0,0,0}\ket{c_1}}%
}}}
\put(1876,-1636){\makebox(0,0)[lb]{\smash{\SetFigFont{10}{12.0}{\familydefault}{\mddefault}{\updefault}{\color[rgb]{0,0,0}$\vdots$}%
}}}
\put(1651,-1186){\makebox(0,0)[lb]{\smash{\SetFigFont{10}{12.0}{\familydefault}{\mddefault}{\updefault}{\color[rgb]{0,0,0}\ket{a_1}}%
}}}
\put(1651,-736){\makebox(0,0)[lb]{\smash{\SetFigFont{10}{12.0}{\familydefault}{\mddefault}{\updefault}{\color[rgb]{0,0,0}\ket{k_n}}%
}}}
\put(1876,-286){\makebox(0,0)[lb]{\smash{\SetFigFont{10}{12.0}{\familydefault}{\mddefault}{\updefault}{\color[rgb]{0,0,0}$\vdots$}%
}}}
\put(1651,164){\makebox(0,0)[lb]{\smash{\SetFigFont{10}{12.0}{\familydefault}{\mddefault}{\updefault}{\color[rgb]{0,0,0}\ket{k_1}}%
}}}
\put(5776,-2461){\rotatebox{90.0}{\makebox(0,0)[lb]{\smash{\SetFigFont{10}{12.0}{\familydefault}{\mddefault}{\updefault}{\color[rgb]{0,0,0}$\bs{l}^T(\bs{r}\cdot\bs{\alpha})\bs{l}$}%
}}}}
\put(6976,-2536){\rotatebox{90.0}{\makebox(0,0)[lb]{\smash{\SetFigFont{10}{12.0}{\familydefault}{\mddefault}{\updefault}{\color[rgb]{0,0,0}$(\bs{l}^T(\bs{r}\cdot\bs{\alpha})\bs{l})^*$}%
}}}}
\put(1651,-4786){\makebox(0,0)[lb]{\smash{\SetFigFont{10}{12.0}{\familydefault}{\mddefault}{\updefault}{\color[rgb]{0,0,0}\ket{c_n}}%
}}}
\put(2326,-1186){\makebox(0,0)[lb]{\smash{\SetFigFont{10}{12.0}{\familydefault}{\mddefault}{\updefault}{\color[rgb]{0,0,0}$X$}%
}}}
\put(4726,-1186){\makebox(0,0)[lb]{\smash{\SetFigFont{10}{12.0}{\familydefault}{\mddefault}{\updefault}{\color[rgb]{0,0,0}$H$}%
}}}
\put(4726,-2086){\makebox(0,0)[lb]{\smash{\SetFigFont{10}{12.0}{\familydefault}{\mddefault}{\updefault}{\color[rgb]{0,0,0}$H$}%
}}}
\put(4126,-2086){\makebox(0,0)[lb]{\smash{\SetFigFont{10}{12.0}{\familydefault}{\mddefault}{\updefault}{\color[rgb]{0,0,0}$X$}%
}}}
\put(2326,-3886){\makebox(0,0)[lb]{\smash{\SetFigFont{10}{12.0}{\familydefault}{\mddefault}{\updefault}{\color[rgb]{0,0,0}${\cal A}$}%
}}}
\end{picture}